\documentclass[11pt, a4paper]{article}
\usepackage[a4paper, left=2.5cm, right=2.5cm, top=2.5cm, bottom=4cm]{geometry}
\usepackage{graphicx} 
\usepackage{appendix}
\usepackage{amsmath}
\usepackage{authblk}
\usepackage{abstract} 
\usepackage[colorlinks=true]{hyperref}
\usepackage{setspace}

\newenvironment{acknowledgements}{\section*{Acknowledgements}}{}

\newif\ifshowemails
\showemailstrue 

\newcommand{\MarceloEmail}{mr.barbosa@unesp.br}

\newcommand{\maybeemail}[1]{%
  \ifshowemails
    \if\relax\detokenize{#1}\relax
    \else
      \thanks{\href{mailto:#1}{\texttt{#1}}}%
    \fi
  \fi
}

\title{Super $T\bar{T}$ deformation and the RNS non-critical superstring}

\author[1,2,3]{Marcelo Rezende Barbosa\maybeemail{\MarceloEmail}}

\affil[1]{ICTP-SAIFR\\ R. Dr. Bento T. Ferraz 271, Bl. II, Sao Paulo 01140-070, SP, Brazil}

\affil[2]{Instituto de F{\'i}sica Te{\'o}rica, UNESP-Universidade Estadual Paulista\\

R. Dr. Bento T. Ferraz 271, Bl. II, Sao Paulo 01140-070, SP, Brazil}
\affil[3]{C.N. Yang Institute for Theoretical Physics,\\

S.U.N.Y. Stony Brook University\\
Stony Brook, NY 11794-3840, USA}

\date{April 2026}

\begin{document}

\maketitle

\begin{abstract}
In this paper we review the super $T\bar{T}$ deformation of $\mathcal{N}=(1,1)$ theories in the superspace formulation, alongside its interpretation in the context of noncritical string theory. By combining the superspace approach with concepts from the study of super-Riemann surfaces, we demonstrate that super $T\bar{T}$ deformations in superspace can be naturally interpreted as a noncritical RNS superstring theory.
We also propose a possible interpretation of the super $T\bar{T}$ deformations as 2D supergravity in the superspace through some field redefinitions.
\end{abstract}

\newpage

\section{Introduction}
Deformations have garnered significant attention within the modern theoretical physics community in recent years. Through the study of deformations, we can explore several aspects of quantum field theories and forge connections between fundamentally different theories.

One deformation that has been extensively studied is the so-called $T\bar{T}$ deformation. We can highlight several aspects of this theory that make it of high interest. The $T\bar{T}$ deformation is defined in a 2D field theory through the operator:
\begin{equation}
    \mathcal{O}_{T\bar{T}} = -\text{det}[T^{\mu}_{\;\;\nu}]=\frac{1}{2}(T^{\mu \nu}T_{\mu \nu}-T^{\mu}_{\;\;\mu}T^{\nu}_{\;\;\nu}).
\end{equation}

One of the primary reasons this deformation is so attractive to theoretical physicists is that it is exactly solvable; meaning that if we know the spectrum of the seed theory, we can derive the spectrum of the deformed theory through the formula \cite{Cavaglia:2016oda,Smirnov:2016lqw,he2025ttbardeformation}: 
\begin{equation}
    E(R,\lambda ) =\frac{R}{2\lambda}\Bigg(1-\sqrt{1-\frac{4\lambda}{R}E_{(0)}+\frac{4\lambda^{2}}{R^{2}}J_{(0)}^{2}}\Bigg), 
\end{equation}
where $E_{(0)}$, and $J_{(0)}$ are respectively, the energy and momentum in the seed theory, $R$, the radius of the cylinder we put the theoy, and $\lambda$, the coupling of the deformation. 

It is important to notice that $E(R,\lambda)=E(\lambda/R)$, so the functional dependence in coupling is fully determined if we now how it changes with the size of the space.

Other reasons why the study of this deformation is compelling include its connections to other theories, such as 2D topological gravity \cite{caputa2020geometrizing,dubovsky2018ttbar,hirano2025ttdeformed}, the Nambu-Goto string theory \cite{he2025ttbardeformation}, and non-critical string theories \cite{Tolley2019,Callebaut2020,Baggio2018}. In this paper, we generalize the relation between the $T\bar{T}$ deformation and string theories to the supersymmetric case.

From the spectrum, it was recognized that the deformation of a scalar theory yields exactly the spectrum expected for a string theory in the Nambu-Goto description in the static gauge \cite{he2025ttbardeformation}. In \cite{Tolley2019}, the authors reverse engineered deformed CFTs by employing the Stückelberg field method to restore conformal symmetry after the deformation. They demonstrated that this procedure yields a Polyakov string theory, in which the spacetime coordinates $X^{0,1}$ are identified with the Stückelberg fields.

In \cite{Callebaut2020}, the opposite path was studied. Starting with a CFT coupled to a Polyakov action for $X^{0,1}$, it was shown that after integrating out the extra fields, the theory is equivalent on-shell to the $T\bar{T}$ deformation of the original CFT. In that paper, they considered two cases: one where the CFT plus the Polyakov action has a vanishing central charge, making it possible to quantize, and another where it has a non-vanishing central charge. In the latter case, it is necessary to couple the theory to 2D gravity on the worldsheet and modify the energy-momentum tensor in a subtle way.

In both cases, only NS-NS fields were turned on, in agreement with \cite{Baggio2018a}, which established the equivalence between integrable deformations and string theories in NS-NS background fields.

In this context, supersymmetry is introduced, and in \cite{Baggio2018} it was shown that the $T\bar{T}$ deformation can be lifted to a description in superspace, thus making it compatible with supersymmetry. The connection between the supersymmetric string in static gauge and $T\bar{T}$ deformations was investigated in \cite{Tseytlin:2025dud}. 

Extensive literature explores the relationship connecting $T\bar{T}$ deformations, supersymmetry, and string theory; notable examples include \cite{Dei:2024sct,Chang:2018dge,Chang:2019kiu}.

In this paper, we bring together the two concepts presented above: the string theory realization of the $T\bar{T}$ deformation and supersymmetry. The paper is organized as follows. In Section 2, we review the basics of the string construction of the $T\bar{T}$ deformation for both critical and non-critical cases. In Section 3, we comment on the lifting of the deformation to the $\mathcal{N}=(1,1)$ 2D superspace, where we briefly review necessary concepts, such as the description of the energy-momentum tensor in the unconstrained formulation of supergravity and the SUSY algebra. In Section 4, we generalize the string construction to the supersymmetric case, using the RNS (or spinning) string in superspace to describe super $T\bar{T}$ deformations. We discuss both critical and non-critical cases, demonstrating that we indeed recover the expected spectrum.

\section{Review of the Bosonic Approach}
In this section, we briefly review the relation between the $T\bar{T}$ deformation and the Polyakov action. We follow closely \cite{Callebaut2020}. We refer to \cite{Callebaut2020,Sabbata} for a more detailed exposition.

We start with a CFT described by some action $S_{CFT}$. We then define the total action as $S_{CFT}$ plus the action of a 2D spacetime string theory in light-cone coordinates, i.e.
\begin{equation}
    S=S_{CFT}+\frac{1}{2\pi}\int d^{2}z\partial v\bar{\partial} u,
\end{equation}
where $v=x^{1}-x^{0}$ and $u=x^{1}+x^{0}$. In order to be quantizable, this CFT should have central charge $c_{CFT}=24$. These fields decompose into holomorphic and anti-holomorphic parts on-shell, i.e.
\begin{equation}
    v(z,\bar{z})=v(\bar{z})+\tilde{v}(z),\,\,\,\,\,\,\,\;\;u(z,\bar{z})=u(z)+\tilde{u}(\bar{z}).
\end{equation}

The space coordinate has the boundary condition
\begin{equation}\label{bosonic b.c}
    x^{1}(e^{2\pi i}z)=x^{1}(z)+2\pi R.
\end{equation}

The Virasoro constraints are
\begin{equation}\label{bosonic virasoro}
    \begin{aligned}
        -\partial u \, \partial v+T_{\mathrm{CFT}}=0, \\ -\bar{\partial} u \, \bar{\partial} v+\bar{T}_{\mathrm{CFT}}=0,
    \end{aligned}
\end{equation}
and we can use an on-shell gauge fixing,
\begin{equation}\label{bosonic gauge fixing}
    \partial u=\bar{\partial} v=1.
\end{equation}

It is important to note that this is a different gauge choice from the one that is usually adopted in the string theory literature. In the standard approach, one typically fixes the variable $v$ (or $u$) completely, rather than only its holomorphic (or anti-holomorphic) part.

We can now insert \eqref{bosonic gauge fixing} into \eqref{bosonic virasoro} and substitute back into the action to obtain the $T\bar{T}$ deformation.

If the theory is a CFT with $C_{CFT}\neq 24$, we need to change the Virasoro constraint in order to obtain an algebra with vanishing total central charge. This amounts to modifying the energy-momentum tensor of the 2D string theory as
\begin{equation}\label{new bosonic e.m}
    \begin{aligned}
       & T=-\partial u \, \partial v-\kappa \, \partial^2 \log \partial u, \\
       & \bar{T}=-\bar{\partial} u \, \bar{\partial} v-\kappa \, \bar{\partial}^2 \log \bar{\partial} v.
    \end{aligned}
\end{equation}

It is not hard to check that \eqref{new bosonic e.m} is a well-defined energy-momentum tensor. We skip this computation here; it can be found in \cite{Callebaut2020}, and a very similar computation will be carried out for the supersymmetric case below.

With this new energy-momentum tensor, the central charge of the 2D string theory is $c=2-24\kappa$, and imposing the vanishing of the total central charge gives
\begin{equation}\label{definition of kappa bosonica}
    \kappa=\frac{c_{CFT}}{24}-1.
\end{equation}

We can again impose the gauge-fixing condition to obtain the correct deformation.

We can now obtain the spectrum of the theory. To do so, we use the Virasoro constraints to derive the level-matching and mass-shell conditions,
\begin{equation}\label{level matching and mass shell}
    \begin{aligned}
        &L_{0}^{\text{total}}+\bar{L}_{0}^{\text{total}}-2=0,\\
        &L_{0}^{\text{total}}-\bar{L}_{0}^{\text{total}}=0.
    \end{aligned}
\end{equation}
We also notice that we can rewrite the action as
\begin{equation}
    S_X=\frac{1}{2 \pi} \int d^2 z\left(G_{a b}+B_{a b}\right) \partial x^a \bar{\partial} x^b,
\end{equation}
for $G_{ab}=\eta_{ab}$ and a constant $B$-field $B_{ab}=\varepsilon_{ab}$.

We also need the mode expansion for $x^a$,
\begin{equation}
    \partial x^a=-\frac{i}{2} \frac{p_L^a}{z}-\frac{i}{\sqrt{2}} \sum_{n \neq 0} \alpha_n^a z^{-n-1}, \quad \bar{\partial} x^a=-\frac{i}{2} \frac{p_R^a}{\bar{z}}-\frac{i}{\sqrt{2}} \sum_{n \neq 0} \bar{\alpha}_n^a \bar{z}^{-n-1},
\end{equation}
where $p_{L,R}^a=\left(p^a \pm\left(G^{a b} \pm B^{a b}\right) w_b\right)$, $p^a$ are spacetime momenta, and $w_a$ are winding zero modes.

We parameterize $p^a$ as
\begin{equation}
    p_L^0=p_R^0=\mathcal{E}+B R, \quad p_L^1=\frac{J}{R}+R, \quad p_R^1=\frac{J}{R}-R,
\end{equation}
where $\mathcal{E}$ is the spacetime energy and $J$ the spacetime angular momentum. Plugging these back into the level-matching and mass-shell conditions, respectively, we get
\begin{equation}\label{level matching and mass shell applyed}
    \begin{aligned}
        &\Delta_{R}-\Delta_L=\frac{1}{4}p_{1L}^{2}-\frac{1}{4}p_{1R}^{2}=J,\\
        &-\frac{1}{2}p_{0}^{2}+\frac{1}{4}p_{1L}^{2}+\frac{1}{4}p_{1R}^{2}+\Delta_{L}+\Delta_{R}-2\kappa -2=0,
    \end{aligned}
\end{equation}
where $\Delta_{L,R}$ are the left- and right-scaling dimensions in the CFT. Substituting \eqref{definition of kappa bosonica} and the parameterizations for $p^{a}$, and writing the energy of the CFT as $E=\frac{1}{R}(\Delta_L+\Delta_R-\frac{c_{CFT}}{12})$, we solve for the spacetime energy $\mathcal{E}$,
\begin{equation}
    \mathcal{E}=R\Bigg(-1+\sqrt{1+\frac{2E}{R}+\frac{J^2}{R^4}}\Bigg).
\end{equation}

We can recognize the spectrum of a $T\bar{T}$-deformed theory when we identify the inverse of the radius as the coupling.

A comment is in order here. The energy-momentum tensor \eqref{new bosonic e.m} can be derived from the action \cite{Callebaut2020}:
\begin{equation}\label{modified action}
    S=\frac{1}{2\pi}\int d^{2}z\bar{\partial}u\partial v-\frac{\kappa}{4}R^{(2)}\,\,\text{ln}(\partial u \bar{\partial} v),
\end{equation}
where $R^{(2)}$ is the 2D curvature scalar on the worldsheet. This action and the energy-momentum tensor are intimately related to 2D gravity, as discussed in \cite{Callebaut2020,Verlinde1993,Chung1993,Cangemi1995}.

We will generalize these arguments for the RNS string using the superspace formulation. Some normalizations will differ from those used in this section, but the calculations and conclusions are the same.

\section{SUSY and \texorpdfstring{$T\bar{T}$}{TbarT}}
In this section we review some basics of 2D supergravity in $\mathcal{N}=(1,1)$ superspace. An incomplete list of references that treat superspace supergravity is \cite{DHoker:1988pdl,deligne1999quantum,grisaru1995quantum,Grisaru1995,Grisaru:1985vx,gates1983superspace,Gates:1985vk,buchbinder1998ideas}.

At the end of this section, we review the uplift of the $T\bar{T}$ deformation to the $\mathcal{N}=(1,1)$ superspace, as carried out in \cite{Baggio2018}.

We start by defining the coordinates in this superspace, $z^M=(z,\bar{z},\theta,\bar{\theta})$, where the first two coordinates are bosonic and the other two are fermionic.
We use unconstrained prepotentials to work with supergravity; that is, we start from a description that solves the supergravity constraints. With that stated, the SUSY operators are
\begin{equation}\label{Q and D}
    \begin{aligned}
        &\mathcal{D}=\frac{\partial}{\partial \theta}+\theta\partial, \;\;\;\;\;\mathcal{Q}=\frac{\partial}{\partial \theta}-\theta\partial,\\
        &\bar{\mathcal{D}}=\frac{\partial}{\partial\bar{\theta}}+\bar{\theta}\bar{\partial},\;\;\;\;\;\bar{\mathcal{Q}}=\frac{\partial}{\partial\bar{\theta}}-\bar{\theta}\bar{\partial}
    \end{aligned}
\end{equation}
and their algebra is
\begin{equation}\label{SUSY algebra}
    \begin{aligned}
        &\mathcal{D}^2=\partial,\;\;\;\;\;\mathcal{Q}^2=-\partial,\\
        &\bar{\mathcal{D}}^2=\bar{\partial},\;\;\;\;\;\bar{\mathcal{Q}}^2=-\bar{\partial}.
    \end{aligned}
\end{equation}
All other anticommutators vanish (equivalently, one may rewrite the above relations using anticommutators and appropriate factors of two).

We have three independent unconstrained prepotentials, which we denote by $H_{-}^{\;z}(z^M)$, $H_{+}^{\;\bar{z}}(z^M)$, and $S(z^M)$. In this notation, $S(z^M)$ is the super-Weyl compensator, and in what follows it decouples from the theory because we work with CFTs.

The coupling between matter and supergravity, in terms of the prepotentials, is
\begin{equation}\label{matter gravityt coupling}
    S=\int d^{2}zd^{2}\theta\bigg(H_{-}^{\;z}\mathcal{T}+H_{+}^{\;\bar{z}}\bar{\mathcal{T}}\bigg).
\end{equation}

Under a superdiffeomorphism, the variations are
\begin{equation}\label{superdiffeo}
    \delta H_{-}^{\;z}=-\bar{\mathcal{D}}\mathcal{K}^{z},\;\;\;\;\;\delta H_{+}^{\;\bar{z}}=-\mathcal{D}\mathcal{K}^{\bar{z}},
\end{equation}
and, using integration by parts and demanding invariance of the action, we obtain the conservation equations
\begin{equation}\label{conservation equations}
    \bar{\mathcal{D}}\mathcal{T}=0,\;\;\;\;\;\mathcal{D}\bar{\mathcal{T}}=0,
\end{equation}
modulo matter equations of motion. 

Focusing on the holomorphic sector and choosing holomorphic coordinates, the super energy-momentum tensor (here denoted by $\mathcal{T}$) has the component decomposition
\begin{equation}
    \mathcal{T}(z,\theta)=G(z)+\theta T(z),
\end{equation}
where $G(z)$ and $T(z)$ are the worldsheet supercurrent and energy-momentum tensor, respectively.

From the above decomposition, it follows straightforwardly that
\begin{equation}\label{T as DmathcalT}
    \mathcal{D}\mathcal{T}=T(z).
\end{equation}
Using the superconservation equations \eqref{conservation equations}, it is also straightforward to derive the usual conservation for $T$, i.e.
\begin{equation}\label{usual conservation}
    \bar{\partial}T=0.
\end{equation}
The same statements hold for the anti-holomorphic part.

All in all, we can define the supersymmetric $T\bar{T}$ operator as
\begin{equation}\label{superttbar operator}
    \mathcal{O}_{sTT}=\bar{\mathcal{T}}\mathcal{T},
\end{equation}
and, working out the SUSY algebra, we can show that it defines an invariant supersymmetric deformation. First, we need to notice that
\begin{equation}\label{ttbar in the superspace}
    T\bar{T}=\mathcal{D}\bar{\mathcal{D}}\mathcal{O}_{sTT}|_{\theta=\bar{\theta}=0}=\int d^2\theta \, \mathcal{O}_{sTT}.
\end{equation}

With the above equation, we can lift the $T\bar{T}$ deformation to a deformation in $\mathcal{N}=(1,1)$ 2D superspace. Then, using standard arguments in superspace calculus, it follows that the superdeformation defined by $\mathcal{O}_{sTT}$ is indeed supersymmetric.

\section{The RNS Superstring}
\subsection{The critical case}
In this section, we connect both the Callebaut et al. formulation using Polyakov strings and the more recently discovered supersymmetric $T\bar{T}$ deformation reviewed in the previous section.
The starting point is the following theory:
\begin{equation}\label{super non critical string}
    S[\Phi,X]=S_{CFT}[\Phi]+\frac{1}{2}\int d\mathbf{\mu}\bar{\mathcal{D}}X_{a}\mathcal{D}X^{a},
\end{equation}
Here, $\Phi$ represents the fields in the CFT defined by the action $S_{CFT}[\Phi]$, and $X_{a=0,1}$ are superscalar fields in a free superscalar theory with target space signature $1+1$. In the above formula, the measure is the measure on the super-Riemann surface with supervielbein determinant $E$: $d\mathbf{\mu}=d^2z\,d^2\theta\, E$.

In the following, we define $\mathcal{T}$ as the super-stress tensor of the pair of super-scalars, $\mathcal{T}_{CFT}$ as the stress tensor for the SCFT, and their conjugates $\bar{\mathcal{T}},\bar{\mathcal{T}}_{CFT}$, respectively.

In what follows, we work with the holomorphic part only, but everything carries over to the anti-holomorphic one.

The super Virasoro constraint that follows from the action \eqref{super non critical string} is
\begin{equation}\label{supervirasoro}
    \mathcal{T}+\mathcal{T}_{CFT}=0,
\end{equation}
and the same holds for the anti-holomorphic currents.

We now define the boundary conditions that will be important in the next steps. In analogy with \eqref{bosonic b.c}, we impose that after a $2\pi$ rotation in the spatial direction,
\begin{equation}\label{susy b.c}
    X^{1}(\text{e}^{2\pi i}z,\text{e}^{-2\pi i}\bar{z},-\theta,-\bar{\theta})=X^{1}(z,\bar{z},\theta,\bar{\theta})+2\pi R.
\end{equation}

From the above equation it follows that the bosonic part of $X^1$ behaves as in the bosonic case, while the fermionic parts have NS--NS boundary conditions. This is in agreement with \cite{Baggio2018a}, which states that the deformations are equivalent to a string in the NS--NS sector.

This conclusion is also important for computing the spectrum, since it depends on the normal-ordering constant, which differs between the sectors of the RNS string.

For what is to come, it is convenient to go to light-cone coordinates:
\begin{equation}
    X^{1}(z,\theta)=\frac{U(z,\theta)+V(z,\theta)}{\sqrt{2}};\;\;\;X^{0}(z,\theta)=\frac{U(z,\theta)-V(z,\theta)}{\sqrt{2}}.
\end{equation}
Here the notation $(z,\theta)$ really means $(z,\bar{z},\theta,\bar{\theta})$.

The action for the pair of scalar fields can be expressed as
\begin{equation}\label{lightconeaction}
    S_{X}=\int d\mu \, \bar{\mathcal{D}}U\mathcal{D}V.
\end{equation}
In these coordinates, the super Virasoro constraints read
\begin{equation}
    \mathcal{T}_{CFT}-\frac{1}{2}[\mathcal{D}U\partial V+\mathcal{D}V\partial U]=0.
\end{equation}

Following Callebaut et al.'s approach for the bosonic case, we now fix the gauge by imposing $\mathcal{D}U=\theta p$, where $p$ is a constant.
This gauge fixing implies a ``light-cone'' type gauge $\partial U=p$. In this setup, we can write
\begin{equation}\label{DVequation}
    \mathcal{D}V=\frac{1}{p}\mathcal{T}_{CFT}+\frac{1}{2}\frac{\partial}{\partial\theta}V.
\end{equation}

Similarly, we can treat the anti-holomorphic equations with the gauge fixing $\bar{\mathcal{D}}V=\bar{\theta}k$, where $k$ is a constant, to get
\begin{equation}\label{DUequation}
    \bar{\mathcal{D}}U=\frac{1}{k}\bar{\mathcal{T}}_{CFT}+\frac{1}{2}\frac{\partial}{\partial\bar{\theta}}U.
\end{equation}

We can now substitute \eqref{DVequation} and \eqref{DUequation} back into the action \eqref{lightconeaction} to get
\begin{equation}\label{superttbargf}
    S[\Phi]=S_{CFT}[\Phi]+\frac{1}{kp}\int d^{2}zd^{2}\theta\,\,\bar{\mathcal{T}}_{CFT}\,\mathcal{T}_{CFT}+\cdots,
\end{equation}
where the $\cdots$ are terms that vanish because of (anti-)holomorphicity and Grassmann integral properties. More precisely, the $\cdots$ are terms proportional to the equations of motion, which vanish on-shell.

Action \eqref{superttbargf} is the action for a $T\bar{T}$-deformed $\mathcal{N}=(1,1)$ SCFT, as we saw in previous sections.

\subsection{Quantization}
To quantize the above theory, we should ensure that the total central charge vanishes. If $c_{CFT}$ is the central charge of the seed theory, the total central charge is
\begin{equation}
    c=c_{CFT}+\frac{3}{2}(2)-15,
\end{equation}
where the $2$ comes from the two super-scalars, and the $-15$ comes from the super-ghosts.

The above result means that, given a super-CFT with central charge $12$, we can interpret its $T\bar{T}$ deformation as a 2D string theory whose internal CFT is given by the seed theory. For a theory of 8 free super-scalars, the $T\bar{T}$ deformation is exactly the 10D string theory, as observed in \cite{Baggio2018} and \cite{Baggio2018a}.

\subsubsection{The Non-Critical Case}
As done in \cite{Callebaut2020}, in the non-critical case we need to modify the stress-energy tensor (the super stress-energy tensor in our case). It turns out that the generalization of the term used in \cite{Callebaut2020} to the supersymmetric case is highly non-trivial. The naive guess would be

\begin{equation}\label{naive guess of deformation}
    \tilde{\mathcal{T}}=-\frac{1}{2}(\mathcal{D}U\partial V+\mathcal{D}V \partial U)-\kappa\mathcal{D}\partial \text{ln}[\partial U],
\end{equation}
but, because $\partial$ is not so well behaved under superconformal transformations (see appendix \ref{app:connections} for a review), this is incorrect. Indeed, as we show below, this deformation is only the first term of the correct deformation.

In the supersymmetric case, the modification that accomplishes this is given by a quasi superprojective connection\footnote{This is also how it works in the bosonic case, but there we did not need to summon such fancy names}:
\begin{equation}\label{new super energy momentum tensor}
    \begin{aligned}
        \\\tilde{\mathcal{T}}=-\frac{1}{2}(\mathcal{D}U\partial V+\mathcal{D}V \partial U)-\kappa\mathcal{D}\partial \text{ln}[\mathcal{D}\Psi_U],
        \\ \bar{\tilde{\mathcal{T}}}=-\frac{1}{2}(\bar{\mathcal{D}}U\bar{\partial} V+\bar{\mathcal{D}}V \bar{\partial} U)-\kappa\bar{\mathcal{D}}\bar{\partial} \text{ln}[\bar{\mathcal{D}}\bar{\Psi}_V],
    \end{aligned}
\end{equation}
where $\Psi_{U}=\frac{\mathcal{D}U}{\sqrt{\partial U}}$, and the anti-holomorphic part is obtained by replacing $\Psi_{U}$ with $\bar{\Psi}_{V}=\frac{\bar{\mathcal{D}}V}{\sqrt{\bar{\partial}V}}$.

To justify the above super stress-energy tensor, we can study the finite transformation of modified stress-energy tensors, i.e. $\tilde{T}=T-\kappa\Phi$, for some composite operator $\Phi$. To do so, we start with the infinitesimal transformation (here we use the bosonic theory, which can be generalized straightforwardly to the supersymmetric one):

\begin{equation}
    \delta \tilde{T}(z)=\oint_{w} \epsilon(w)\tilde{T}(w)\tilde{T}(z),
\end{equation}
where we assume radial ordering and use the normalization $\oint_{w}=\frac{1}{2\pi i}\oint dw$.
From this, we can derive
\begin{equation}\label{infinitesimal transformation of modified emt}
    \delta\tilde{T}=-\frac{1}{12}c\partial^{3}\epsilon-2T\partial\epsilon-\epsilon\partial T+a[u;z,z^\prime]+\kappa\delta^{\prime}\Phi.
\end{equation}

In \eqref{infinitesimal transformation of modified emt}, $a[u;z,z^{\prime}]$ is a field-($u$)-dependent infinitesimal anomaly, and $\delta^{\prime}\Phi$ is the variation generated by the original stress-energy tensor $T$. These quantities are given by
\begin{equation}\label{definition of anomaly and phi var}
    \begin{aligned}
        \\a[u;z,z^{\prime}]\equiv-\kappa\oint_{w}\epsilon(w)\Phi(w)T(z),\\ \kappa\delta^{\prime}\Phi\equiv-\kappa\oint_{w}\epsilon(w)T(w)\Phi(z).
    \end{aligned}
\end{equation}
We assume (as in the case of interest) that the OPE $\Phi\Phi$ has only regular terms.

In order for $\tilde{T}$ to be an allowed stress-energy tensor, the variation \eqref{infinitesimal transformation of modified emt} should take the form
\begin{equation}\label{allowed emt}
    \delta \tilde{T}=-\frac{1}{12}\tilde{c}\partial^{3}\epsilon-2\tilde{T}\partial\epsilon-\epsilon\partial \tilde{T},
\end{equation}
which implies
\begin{equation}
    a[u;z,z^{\prime}]+\kappa\delta^{\prime}\Phi=-\frac{1}{12}c^{\prime}\partial^{3}\epsilon+2(\kappa \Phi)\partial\epsilon+\epsilon\partial (\kappa\Phi),
\end{equation}
where $c^{\prime}=c^{\prime}(\kappa)$ and $\tilde{c}=c+c^{\prime}$. Following \cite{Callebaut2020} closely, we say that $a[u;z,z^{\prime}]$ can be separated into a field-dependent anomaly and a central-charge anomaly term, and that $\delta^{\prime}\Phi$ transforms almost as a stress-energy tensor, differing from it by a field-dependent anomaly. The field-dependent anomalies of both contributions must cancel, and the central-charge anomalies must add up to give $c^{\prime}$.

Now, integrating the variation of $\Phi$ (with appropriate normalization) gives
\begin{equation}\label{anomaly equation1}
    \Phi^{\prime}{dz^{\prime}}^{2}=\Phi dz^{2}+\frac{1}{12}\{f(z),z\}dz^{2}- \mathcal{A}[u],
\end{equation}
where $\mathcal{A}$ is the integrated anomaly, properly normalized. This is the point we need in order to motivate the form of \eqref{new super energy momentum tensor}.

When $\mathcal{A}=0$, $\Phi$ is a well-known quantity, called a projective affine connection (see appendix \ref{app:connections} and \cite{Ader1993,Ader1992,ader1992polyakov,GIERES1993}), and is given by
\begin{equation}
    \Phi(u)=\partial\gamma(u)-\frac{1}{2}[\gamma(u)]^{2},\,\,\,\,\gamma(u)=\partial\text{ln}(\partial u).
\end{equation}
In the above equation, $\gamma$ is an affine connection.

We can recognize from the above equation that the first term is indeed the modification used in \cite{Callebaut2020}.

We can generalize the transformations and objects mentioned above to $\mathcal{N}=(1,1)$ superspace. The super affine connection is $\Gamma(U)=\mathcal{D}\text{ln}(\mathcal{D}\Psi_{U})$, where $\mathcal{D}U=\Psi_{U}\mathcal{D}\Psi_{U}$. As commented in appendix \ref{app: psiU}, this can be solved for $\Psi_{U}$, giving $\Psi_{U}=\frac{\mathcal{D}U}{\sqrt{\partial U}}$. The super projective connection is
\begin{equation}
    \Phi(U)=-[\partial\Gamma(U)+\Gamma(U)\mathcal{D}\Gamma(U)],
\end{equation}
and again we can recognize the first term as the proposed modification for the supersymmetric case.

This modification indeed reproduces a new allowed super stress-energy tensor, as can be checked below.

We can write \eqref{new super energy momentum tensor} in terms of derivatives of $U$ only:
\begin{equation}\label{new emt in terms of U}
    \tilde{\mathcal{T}}=\mathcal{T}-\kappa\frac{1}{2}\Bigg[\mathcal{D}\partial\text{ln}(\partial U)+\mathcal{D}\partial   \bigg[\frac{\mathcal{D}U\mathcal{D}\partial U}{(\partial U)^{2}} \bigg]\Bigg].
\end{equation}

We can now compute the OPEs for this new stress-energy tensor. Using the notation $Z_{i}=(z_{i},\theta_{i})$, $Z_{ij}=z_{i}-z_{j}-\theta_{i}\theta_{j}$, and $\theta_{ij}=\theta_{i}-\theta_{j}$, we get
\begin{equation}\label{ope of new energy momentum tensor}
    \tilde{\mathcal{T}}(Z_{1})\tilde{\mathcal{T}}(Z_{2})\sim \frac{\frac{1}{4}\hat{\tilde{c}}}{Z_{12}^{3}}+\frac{3}{2}\frac{\theta_{12}}{Z_{12}^{2}}\tilde{\mathcal{T}}(Z_{2})+\frac{1}{2}\frac{1}{Z_{12}}\mathcal{D}\tilde{\mathcal{T}}(Z_{2})+\frac{\theta_{12}}{Z_{12}}\partial \tilde{\mathcal{T}}(Z_{2}),
\end{equation}
where $\tilde{c}=\frac{3}{2}(2)-12\kappa=3-12\kappa$. If we now impose the vanishing of the total central charge (original CFT + 2D string + ghosts), we get
\begin{equation}\label{definition of kappa}
    0=c_{CFT}+(3-12\kappa)-15\implies\kappa=\frac{c_{CFT}}{12}-1.
\end{equation}

We can similarly work out the anti-holomorphic part.

To end this section, we compute the useful OPEs whose main result is \eqref{ope of new energy momentum tensor}. We focus on the holomorphic part only, since the anti-holomorphic part follows in the same way.

The starting point is the OPE between the fields $U$ and $V$:
\begin{equation}
    U(Z_{1})V(Z_{2})=V(Z_{1})U(Z_{2})\sim-\text{ln}Z_{12}.
\end{equation}

The next step is to compute the OPE between $V$ and
\begin{equation}
    \mathcal{D}\Psi_{U}=\sqrt{\partial U}+\frac{1}{2}\mathcal{D}U\mathcal{D}\partial U\frac{1}{(\partial U)^{\frac{3}{2}}},
\end{equation}
using the OPE above.

Then, using the chain rule and performing the calculations, we get
\begin{equation}
    \begin{aligned}
        V(Z_1) \mathcal{D} \Psi_U(Z_2) \sim &\frac{1}{2Z_{12}} (\partial U(Z_{2}))^{-1/2} + \frac{\theta_{12}}{2Z_{12}^2} (\mathcal{D}U(Z_{2}))(\partial U(Z_{2}))^{-3/2}\\
        & - \frac{\theta_{12}}{2Z_{12}} (\partial U(Z_{2}))^{-3/2} \,\partial \mathcal{D}U(Z_{2}) - \frac{3}{4Z_{12}} (\mathcal{D}U(Z_{2}))(\partial U(Z_{2}))^{-5/2} \,\partial \mathcal{D}U(Z_{2}).
    \end{aligned}
\end{equation}

Using the above OPE, we can carefully apply $\mathcal{D}$ and $\partial$ in the first variable, $Z_{1}$, to obtain
\begin{equation}\label{primary 1/2 dim}
    \mathcal{T}(Z_{1})\mathcal{D}\Psi_U(Z_{2})\sim \frac{1}{2}\frac{\theta_{12}}{Z_{12}^{2}}\mathcal{D}\Psi_{U}(Z_{2})+\frac{1}{2}\frac{1}{Z_{12}}\mathcal{D}(\mathcal{D}\Psi_{U}(Z_{2}))+\frac{\theta_{12}}{Z_{12}}\partial(\mathcal{D}\Psi_{U}(Z_{2})).
\end{equation}

From the above OPE, we see that $\mathcal{D}\Psi_{U}$ is a superprimary of conformal dimension $h=1/2$.

To simplify notation in the next steps, we set $J=\mathcal{D}\Psi_{U}$. Using the above OPE and the chain rule, we obtain
\begin{equation}
    \mathcal{T}(Z_{1})\text{ln}[J(Z_{2})]\sim \frac{1}{2}\frac{\theta_{12}}{Z_{12}^{2}}+\frac{1}{2}\frac{1}{Z_{12}}\mathcal{D}(\text{ln}[J(Z_{2})])+\frac{\theta_{12}}{Z_{12}}\partial(\text{ln}[J(Z_{2})]).
\end{equation}

Now we apply $\mathcal{D}_{2}\partial_{2}$ to the above OPE, being careful with Grassmann properties of $\mathcal{T}$ and $\mathcal{D}$:

\begin{equation}\label{tddln}
    \begin{aligned}
       \mathcal{T}(Z_1) \mathcal{D}_2\partial_2 \ln J(Z_2) \sim & \frac{1}{Z_{12}^3} + \frac{\theta_{12}}{Z_{12}^3}\mathcal{D}_2\ln J(Z_2) + \frac{1}{2Z_{12}^2}\partial_2\ln J(Z_2)  \\
       & + \frac{3\theta_{12}}{2Z_{12}^2}\mathcal{D}_2\partial_2\ln J(Z_2) + \frac{1}{2Z_{12}}\partial_2^2\ln J(Z_2)\\
       & + \frac{\theta_{12}}{Z_{12}}\mathcal{D}_2\partial_2^2\ln J(Z_2).
    \end{aligned}
\end{equation}

Finally, to derive \eqref{ope of new energy momentum tensor}, we need to compute $(\mathcal{D}\partial \text{ln}J(Z_{1})) \mathcal{T}(Z_{2})$. The simplest way to perform this calculation is to swap $1\leftrightarrow2$ in \eqref{tddln}, anticommute the operators (paying attention to Grassmann properties), and expand the functions of $Z_{1}$ appearing on the right-hand side. This gives
\begin{equation}\label{ddlnt}
    (\mathcal{D}\partial \ln J(Z_{1})) \mathcal{T}(Z_{2}) \sim \frac{1}{Z_{12}^3} - \frac{\theta_{12}}{Z_{12}^3}\mathcal{D}\ln J(Z_{2}) - \frac{1}{2Z_{12}^2}\partial\ln J(Z_{2}).
\end{equation}

Adding \eqref{ddlnt} and \eqref{tddln}, we obtain
\begin{equation}\label{resultope}
    \begin{aligned}
        \sim & \frac{2}{Z_{12}^3} + \frac{3\theta_{12}}{2Z_{12}^2}\mathcal{D}_2\partial_2\ln J(Z_2) + \frac{1}{2Z_{12}}\mathcal{D}_{2}(\mathcal{D}_{2}\partial_2\ln J(Z_2))\\
        & + \frac{\theta_{12}}{Z_{12}}\partial_{2}(\mathcal{D}_2\partial_2\ln J(Z_2)).
    \end{aligned}
\end{equation}

Adding this to the OPE of $\mathcal{T}\mathcal{T}$ yields \eqref{ope of new energy momentum tensor}.

\section{Spectrum}
To compute the spectrum, we follow again \cite{Callebaut2020}. The (super)Virasoro constraint is
\begin{equation}
    \mathcal{T}_{CFT}-\frac{1}{2}(\mathcal{D}U\partial V+\mathcal{D}V \partial U)+\kappa\frac{1}{2}\Bigg[\mathcal{D}\partial\text{ln}(\partial U)+\mathcal{D}\partial   \bigg[\frac{\mathcal{D}U\mathcal{D}\partial U}{(\partial U)^{2}} \bigg]\Bigg]=0.
\end{equation}
We can proceed exactly as described in the review section, but being careful with some subtleties that appear in the computation.

The first subtlety is that, for the NS sector of the RNS string, we have $L_{0}^{\text{total}}+\bar{L}_{0}^{\text{total}}-1=0$, instead of $L_{0}^{\text{total}}+\bar{L}_{0}^{\text{total}}-2=0$. As the ``2'' in the bosonic case is crucial for the correct spectrum, the ``1'' in the supersymmetric case is crucial as well.

The second subtlety appears when we try to naively use the on-shell conditions of the previous paragraph to compute the spectrum. At first sight, it seems that we have $T\bar{T}$ deformations plus extra contributions coming from the fermions. This can be seen by separating the components of the new super stress-energy tensor:
\begin{equation}
    \tilde{\mathcal{T}}=\tilde{G}(z)+\theta\tilde{T}(z),
\end{equation}
which gives
\begin{equation}\label{components of superT}
    \begin{aligned}
        &\tilde{T}(z)=T(z)-\frac{\kappa}{2}\Bigg\{ \partial^2 \ln(\partial u) + (\partial u)^{-2} \Bigg[ 2\left( \partial^2\ln(\partial u) - 2(\partial\ln(\partial u))^2 \right) (\partial\psi_{u})\psi_{u}  \\
        &\quad + 4\partial\ln(\partial u) (\partial^2\psi_{u})\psi_{u} - (\partial^3\psi_{u})\psi_{u} - (\partial^2\psi_{u})(\partial\psi_{u}) \Bigg] \Bigg\}\\
        &\tilde{G}(z)=G(z)-\frac{\kappa}{2} (\partial u)^{-1} \left[ 2\partial^2\psi_{u} - 3 \partial\ln(\partial u) \partial\psi_{u} - \left( \partial^2 \ln(\partial u) - (\partial \ln(\partial u))^2 \right) \psi_{u} \right].
    \end{aligned}
\end{equation}
The fermionic terms in the improved stress tensor do not
contribute to its zero mode. To see this, define
\begin{equation}
    F(z) \equiv
    \frac{\psi_u(z)\,\partial\psi_u(z)}{(\partial u(z))^2}.
\end{equation}
Using the Grassmann parity of $\psi_u$, direct differentiation
gives
\begin{align}
    \partial^2 F
    = \frac{1}{(\partial u)^2}
    \Big[
    &2\big(
        \partial^2\ln(\partial u)
        -2(\partial\ln(\partial u))^2
    \big)(\partial\psi_u)\psi_u
    \nonumber\\
    &+4\partial\ln(\partial u)
        (\partial^2\psi_u)\psi_u
    -(\partial^3\psi_u)\psi_u
    -(\partial^2\psi_u)(\partial\psi_u)
    \Big].
\end{align}
Thus the component expression above can be written as
\begin{equation}
    \widetilde T(z)
    =
    T(z)
    -\frac{\kappa}{2}\partial^2\ln(\partial u(z))
    -\frac{\kappa}{2}\partial^2 F(z).
    \label{eq:improved-stress-total-derivative}
\end{equation}
These identities also hold for the normally ordered
operators: the $u$--$u$, $\psi_u$--$\psi_u$, and
$u$--$\psi_u$ OPEs are regular, so rearranging these fields
does not generate contraction terms.

We work in the NS sector and in a nonzero longitudinal
momentum sector, defining inverse powers of $\partial u$
by their expansion about its nonzero zero-mode background.
The composite $F$ is single-valued around the quantization
contour: its two NS fermion factors have cancelling
monodromies. Consequently,
\begin{equation}
    \oint\frac{dz}{2\pi i}\,z\,\partial^2 F(z)
    =
    \oint\frac{dz}{2\pi i}\,
    \partial\big[z\,\partial F(z)-F(z)\big]
    =0.
    \label{eq:fermionic-improvement-zero-mode}
\end{equation}
The fermionic improvement therefore makes no contribution
to $\widetilde L_0$. This is an operator identity in the
specified sector and does not require a fermion-number
selection rule on physical states.

The logarithmic bosonic term must be treated separately
because of its monodromy. Writing
\begin{equation}
    \partial u(z)
    =
    \frac{a}{z}\big[1+A(z)\big],
    \qquad a\neq0,
\end{equation}
where $A(z)$ contains the nonzero longitudinal modes, we have
\begin{equation}
    \ln(\partial u)
    =
    \ln a-\ln z+\ln(1+A).
\end{equation}
Here $\ln(1+A)$ is defined by its formal oscillator
expansion and is single-valued. Its second derivative
therefore has vanishing contribution to the zero-mode
contour integral by the same argument, whereas
$\partial^2(-\ln z)=z^{-2}$. Hence
\begin{equation}
    \widetilde L_0=L_0-\frac{\kappa}{2}.
\end{equation}
The antiholomorphic sector gives the identical shift,
with the corresponding nonzero longitudinal zero mode.
It follows that
\begin{equation}
    \widetilde L_0+\widetilde{\bar L}_0
    =
    L_0+\bar L_0-\kappa.
\end{equation}
Thus the improvements contribute precisely the constant
shift $-\kappa$ to the sum of the zero-mode constraints.
Using the NS intercept and proceeding with the momentum
expansion as in the bosonic case, we obtain

\begin{equation}
    L_0^{\text {total}}+\bar{L}_0^{\text {total}}-1=-\frac{1}{2} p_0^2+\frac{1}{4} p_{1 L}^2+\frac{1}{4} p_{1 R}^2+\Delta_L+\Delta_R- \kappa-1=0,
\end{equation}
and using \eqref{definition of kappa} we obtain
\begin{equation}\label{def of the spectrum}
    -2\left(\frac{\mathcal{E}}{2}+\frac{B R}{2}\right)^2+\frac{J^2}{2 R^2}+\frac{R^2}{2}+R E=0.
\end{equation}
Solving for $\mathcal{E}$, we recover the correct $T\bar{T}$ spectrum:
\begin{equation}
    \mathcal{E}=R\Bigg(-1+\sqrt{1+\frac{2E}{R}+\frac{J^2}{R^4}}\Bigg).
\end{equation}

This agrees with the expectation for this deformation.\footnote{The argument presented above is new. The argument in the previous versions is incorrect.}

\section{Supergravity and Modified action}

Last but not least, we can try to understand the action that gives rise to \eqref{new super energy momentum tensor}. The action whose stress-energy tensor is \eqref{new super energy momentum tensor} is
\begin{equation}\label{modified super action}
    S=\int d\mu \, \bigg(\bar{\mathcal{D}}U\mathcal{D}V-\kappa^{\prime} \mathcal{R}\,\, \text{ln}(\mathcal{D}\Psi_{U}\bar{\mathcal{D}}\bar{\Psi}_{V})\bigg),
\end{equation}
where $\mathcal{R}$ is the supercurvature scalar.

That said, we can interpret the term with $\varrho(U,V)=\text{ln}(\mathcal{D}\Psi_{U}\bar{\mathcal{D}}\bar{\Psi}_{V})$ as a super-dilaton-type coupling. Furthermore, defining $\Omega=\Psi_{U}\bar{\Psi}_V$, we can rewrite the modified action as the supergravity action:
\begin{equation}\label{supergravity}
    S=\int d\mu \, \bigg(\Omega\text{e}^\varrho-\kappa^{\prime} \mathcal{R}\,\, \varrho\bigg).
\end{equation}

The above action seems to be related to 2D supergravity in superspace, and it opens discussion for a possible interpretation of the super $T\bar{T}$ deformation as being equivalent to a coupling to supergravity.

\section{Conclusion and Next Steps}
As shown in the main text, the super $T\bar{T}$ deformation of a CFT can be recast as a non-critical spinning string theory (or RNS string theory) in a two-dimensional spacetime upon fixing the light-cone gauge. Our construction generalizes the method presented in \cite{Callebaut2020} to the supersymmetric framework of \cite{Baggio2018}. In particular, it provides an RNS realization of the Green--Schwarz argument developed in \cite{Baggio2018}, where the deformation of eight superscalars was shown to be equivalent to a superstring theory. As in the bosonic case, our approach is more general, since it can, in principle, be applied to any supersymmetric field theory formulated in two-dimensional $\mathcal{N}=(1,1)$ superspace.

In this work, we have focused on extending the relation between the $T\bar{T}$ deformation and Polyakov string theory (or its RNS counterpart in the supersymmetric case). However, our analysis does not include a covariant quantization of the resulting theory, as was carried out initially in \cite{Callebaut2020} and subsequently studied in greater detail in \cite{Sabbata}.

Another important question raised by our construction concerns the connection between the super $T\bar{T}$ deformation and (topological) supergravity in superspace. Such a relation may emerge through suitable field redefinitions of the target-space fields.

Two additional directions for future investigation can be identified, one closely related to the previous discussion and the other motivated by holography. First, given the possibility of performing a covariant quantization in the RNS formalism, it is natural to ask whether super $T\bar{T}$ deformations admit a pure-spinor formulation.

The second direction concerns holography. Can this interpretation of the $T\bar{T}$ deformation provide new insight into a top-down approach to the holographic duality of deformed CFTs?

\begin{acknowledgements}
The author would like to thank Horatiu Nastase for fruitful discussions and valuable corrections on the manuscript, and Martin Rocek for his useful insights and guidance during the project. The author would also like to thank Joao Pedro Gomide for careful review and suggestions, and Arkady Tseytlin for bringing to the author's attention his paper \cite{Tseytlin:2025dud}. The author also gratefully acknowledges the warm hospitality of the Yang Institute for Theoretical Physics (YITP) at Stony Brook University, where this work was written. During the preparation of this work, the author used the Gemini AI model to assist with manuscript drafting and the verification of selected calculations. This research was supported by the São Paulo Research Foundation (FAPESP) under grant No. 2022/05152-2.
\end{acknowledgements}

\begin{appendices}
    \section{Solution of \texorpdfstring{$\Psi_U$}{PsiU}}\label{app: psiU}
    The defining equation for $\Psi_{U}$($=\Psi$ to abreviate notation) is 
    \begin{equation}
        \mathcal{D}U=\Psi\mathcal{D}\Psi.
    \end{equation}
    From the above equation, multiplying $\Psi$ from the left, we get 
    \begin{equation}
        \Psi\mathcal{D}U=0\implies\Psi\propto\mathcal{D}U.  
    \end{equation}
    Putting the ansatz $\Psi =f\mathcal{D}U$, back in the defining equation, we get $f=\frac{1}{\sqrt{\partial U}}$.

    \section{Review of Superprojective Affine Connections}
\label{app:connections}

This appendix provides a brief review of projective and affine connections, starting with the standard bosonic theory on Riemann surfaces and subsequently generalizing to the supersymmetric (SUSY) framework on super Riemann surfaces (SRS). The focus is on their transformation properties and the fundamental relationships connecting them.

\subsection{The Bosonic Case}

\subsubsection{Conformal Transformations and the Schwarzian Derivative}
On a Riemann surface equipped with local holomorphic coordinates $(z, \bar{z})$, a change of local coordinates to another system $(z', \bar{z}')$ is given by a conformal (biholomorphic) transformation:
\begin{equation}
    z' = z'(z), \quad \bar{z}' = \bar{z}'(\bar{z})
\end{equation}
For any such smooth conformal change of coordinates, the \textit{Schwarzian derivative} $S(z'; z)$ of the transformation is defined as:
\begin{equation}
    S(z'; z) \equiv \partial^2 \ln(\partial z') - \frac{1}{2} \left( \partial \ln(\partial z') \right)^2
\end{equation}
where $\partial \equiv \frac{\partial}{\partial z}$. The Schwarzian derivative vanishes if and only if the mapping $z \to z'(z)$ is a projective (Möbius) transformation.

\subsubsection{Projective and Affine Connections}
A \textit{projective connection} (or Schwarzian connection) is a collection of locally holomorphic functions $\{R_{zz}(z)\}$ that do not transform strictly as tensors, but instead pick up an inhomogeneous term proportional to the Schwarzian derivative under conformal mappings:
\begin{equation}
    R_{z'z'}(z') = (\partial z')^{-2} \left[ R_{zz}(z) - S(z'; z) \right]
\end{equation}

Similarly, an \textit{affine connection} is a collection of locally holomorphic functions $\{\gamma_z(z)\}$ that transform with an inhomogeneous logarithmic derivative under coordinate changes:
\begin{equation}
    \gamma_{z'}(z') = (\partial z')^{-1} \left[ \gamma_z(z) - \partial \ln(\partial z') \right]
\end{equation}

\subsubsection{Relation Between Affine and Projective Connections}
Given an affine connection $\gamma_z$, one can always construct a projective connection $R_{zz}$. The relationship between the two is expressed through a Riccati-type equation (often referred to as a Miura transformation in the context of integrable systems):
\begin{equation}
    R_{zz} = \partial_z \gamma_z - \frac{1}{2} \gamma_z^2
\end{equation}
This relation ensures that the resulting quantity $R_{zz}$ satisfies the correct transformation law involving the Schwarzian derivative.

\subsection{The Supersymmetric Case}

\subsubsection{Superconformal Transformations and the Super Schwarzian}
In the supersymmetric setting, a super Riemann surface (SRS) is locally parametrized by superconformal coordinates $(z, \bar{z}, \theta, \bar{\theta})$, where $\theta$ and $\bar{\theta}$ are odd Grassmann variables. A superconformal change of coordinates $(z, \theta) \to (z', \theta')$ is a smooth mapping satisfying the \textit{superconformal condition}:
\begin{equation}
    \mathcal{D} z' = \theta' \mathcal{D} \theta'
\end{equation}
where $\mathcal{D}$ is the superderivative defined by $\mathcal{D} \equiv \frac{\partial}{\partial \theta} + \theta \frac{\partial}{\partial z}$.

To describe the transformation laws, it is convenient to define an even superfield $w$ via the relation $e^{-w} \equiv \mathcal{D} \theta'$. The \textit{super Schwarzian derivative} $\mathcal{S}(z', \theta'; z, \theta)$ associated with the superconformal map is then defined by:
\begin{equation}
    \mathcal{S}(z', \theta'; z, \theta) \equiv - \left[ \partial \mathcal{D} w + (\partial w)(\mathcal{D} w) \right]
\end{equation}
The super Schwarzian derivative properly generalizes its bosonic counterpart, vanishing identically for superprojective transformations.

\subsubsection{Super Projective and Super Affine Connections}
A \textit{super projective connection} is a collection of odd, locally superanalytic superfields $\{\mathcal{R}_{z\theta}(z, \theta)\}$ that transform under superconformal mappings as:
\begin{equation}
    \mathcal{R}_{z'\theta'}(z', \theta') = e^{3w} \left[ \mathcal{R}_{z\theta}(z, \theta) - \mathcal{S}(z', \theta'; z, \theta) \right]
\end{equation}

By analogy, a \textit{super affine connection} is a collection of odd superfields $\{\Gamma_\theta(z, \theta)\}$ defining a covariant derivative $\nabla \equiv \mathcal{D} + p\Gamma_\theta$. It transforms under superconformal coordinate changes according to:
\begin{equation}
    \Gamma_{\theta'}(z', \theta') = e^{w} \left[ \Gamma_\theta(z, \theta) - \mathcal{D} w \right]
\end{equation}

\subsubsection{Relation Between Super Affine and Super Projective Connections}
Just as in the bosonic theory, a super projective connection $\mathcal{R}_{z\theta}$ can be constructed from a super affine connection $\Gamma_\theta$. This relationship is governed by the super Miura transformation:
\begin{equation}
    \mathcal{R}_{z\theta} \equiv - \left[ \partial \Gamma_\theta + \Gamma_\theta (\mathcal{D} \Gamma_\theta) \right]
\end{equation}
This fundamental identity guarantees that the combination on the right-hand side perfectly absorbs the inhomogeneous shift of $\Gamma_\theta$ to produce the correct super Schwarzian transformation law required for $\mathcal{R}_{z\theta}$.

\end{appendices}

\nocite{*}
\bibliographystyle{unsrt} 
\bibliography{bepe} 

\end{document}